\documentclass[aps,prl,twocolumn,superscriptaddress,showpacs]{revtex4}
\usepackage{bm}
\usepackage{epsf}
\usepackage{amssymb}
\usepackage{amsmath}
\usepackage{graphicx}
\usepackage{rotating}
\usepackage{epsfig}
\usepackage{psfrag}
\usepackage{amsmath}
\usepackage[breaklinks]{hyperref}
\usepackage{hyperref}

\hypersetup{
    bookmarks=true,         
    unicode=false,          
    pdftoolbar=true,        
    pdfmenubar=true,        
    pdffitwindow=true,      
    pdftitle={My title},    
    pdfauthor={Author},     
    pdfsubject={Subject},   
    pdfcreator={Creator},   
    pdfproducer={Producer}, 
    pdfkeywords={keywords}, 
    pdfnewwindow=true,      
    colorlinks=true,       
    linkcolor=red,          
    citecolor=blue,        
    filecolor=magenta,      
    urlcolor=blue           
}

\DeclareMathAlphabet{\bi}{OML}{cmm}{b}{it}

\begin{document}
\def\G{{\cal G}}
\def\F{{\cal F}}
\def\ea{\textit{et al.}}
\def\bM{{\bm M}}
\def\bN{{\bm N}}
\def\bV{{\bm V}}
\def\bj{\bm{j}}
\def\bSig{{\bm \Sigma}}
\def\bLam{{\bm \Lambda}}
\def\bfeta{{\bf \eta}}
\def\bn{{\bf n}}
\def\d{{\bf d}}
\def \xy{$x$--$y$ }
\def\bP{{\bf P}}
\def\bK{{\bf K}}
\def\bk{{\bf k}}
\def\bkn{{\bf k}_{0}}
\def\bx{{\bf x}}
\def\bz{{\bf z}}
\def\bR{{\bf R}}
\def\br{{\bf r}}
\def\bu{{\bm u}}
\def\bq{{\bf q}}
\def\bp{{\bf p}}
\def\by{{\bf y}}
\def\bQ{{\bf Q}}
\def\bs{{\bf s}}
\def\bA{{\mathbf A}}
\def\bv{{\bf v}}
\def\b0{{\bf 0}}
\def\la{\langle}
\def\ra{\rangle}
\def\Im{\mathrm {Im}\;}
\def\Re{\mathrm {Re}\;}
\def\beq{\begin{equation}}
\def\eeq{\end{equation}}
\def\bdm{\begin{displaymath}}
\def\edm{\end{displaymath}}
\def\bnab{{\bm \nabla}}
\def\Tr{{\mathrm{Tr}}}
\def\bJ{{\bf J}}
\def\bU{{\bf U}}
\def\bPsi{{\bm \deltaDelta}}
\def\mA {\mathrm{A}}
\def \R{R_{\mathrm{s}}}
\def \rhos{n_{\mathrm{s}}}
\def \rhon{\tilde{n}}
\def \Rd{R_{\mathrm{d}}}
\def \xy{three dimensional $XY\;$}
\def\sfrac{\textstyle\frac}
\def\e0{\epsilon_B}
\def\ath{a_{3}}
\def\atw{a_{2}}
\def\ntw{n_{\mathrm{2D}}}
\def\2d{2d}
\def\3d{3d}
\def\gtw{\ln(k_F\atw)}
\def\gth{-1/(k_F\ath)}

\title{Apparent low-energy scale invariance in two-dimensional Fermi gases}
\author{Edward~Taylor}
\affiliation{Department of Physics and Astronomy, McMaster University, Hamilton, Ontario, L8S 4M1, Canada}
\author{Mohit~Randeria}
\affiliation{Department of Physics, The Ohio State University, Columbus, Ohio, 43210, USA}

\date{Sep. 27, 2012}

\begin{abstract}
Recent experiments on a $\2d$ Fermi gas find an undamped breathing mode 
at twice the trap frequency over a wide range of parameters. 
To understand this seemingly scale-invariant behavior in a 
system with a scale, we derive two exact results valid across the entire BCS-BEC crossover 
at all temperatures. First, we relate the shift of the mode frequency from its 
scale-invariant value to $\gamma_d \equiv (1+2/d)P-\rho(\partial P/\partial\rho)_s$ in $d$ dimensions. Next, we
relate $\gamma_d$ to dissipation via a new low-energy bulk viscosity sum rule. We argue that $\2d$ is special, 
with its logarithmic dependence of the interaction on density,  and thus $\gamma_2$ is 
small in both the BCS and BEC regimes, even though $P\!-\!2\varepsilon/d$,
sensitive to the dimer binding energy that breaks scale invariance, is not.
\end{abstract}

\maketitle
Systems exhibiting scaling symmetry or conformal invariance are very special. In all laboratory realizations,
one needs to tune one or more physical parameters (temperature, chemical potential, coupling) to observe
scale-invariant behavior, for instance, in the vicinity of a quantum critical point~\cite{Sachdevbook}. 
Another example is provided by strongly interacting Fermi gases in three spatial dimensions ($\3d$), which display remarkable scale invariance properties 
at unitarity, where the $s$-wave scattering length diverges by tuning to the Feshbach resonance. 
This is manifested in universal thermodynamics~\cite{Ho04}, the vanishing of the d.c. bulk viscosity~\cite{Son07}, 
and the entire bulk viscosity spectral function $\zeta(\omega,T)$~\cite{Taylor10} at unitarity.
There may also be tantalizing connections between the ratio of the shear viscosity $\eta$ to the entropy density $s$
of the unitary Fermi gas~\cite{Schafer09,Enss11} and 
the bound for $\eta/s$ conjectured on the basis of gauge/gravity duality~\cite{N4SSYM}.

For the unitary gas in a $\3d$ isotropic harmonic trap, scale invariance manifests itself most dramatically as 
an undamped monopole breathing mode oscillating at twice the trap frequency $\omega_0$
independent of temperature~\cite{Castin04,Taylor09}. 
This mode corresponds to an isotropic \emph{dilation} of the gas wherein the coordinates in the many-body 
wavefunction are scaled $\propto\cos(\omega t)$.  
Scale invariance implies that this wavefunction is an exact eigenstate of the Hamiltonian and oscillates at a 
frequency $2\omega_0$ without damping~\cite{Castin04}.  

In a recent experiment~\cite{Vogt12}, collective modes in a two-dimensional ($\2d$) Fermi gas
were measured over a broad range of temperatures and interaction strengths.
Remarkably, the breathing mode was found to oscillate without any observable damping
at $\simeq 2\omega_0$ for $0.37 \lesssim T/T_F \lesssim 0.9$
and $0\lesssim \ln(k_F\atw) \lesssim 500$, where $a_2$ is the $\2d$ scattering length. This observation is extremely surprising, given that
there is no \textit{a priori} reason to expect scale-invariant behavior in a system which has
a scale, namely, the dimer binding energy in $\2d$. 

Our goal is to understand why the $\2d$ Fermi gas appears to show nearly scale invariant
behavior over a very broad range of parameters without the need for fine-tuning.  
Understanding this may give insight into related problems such as why, in some quantum field theories with 
conformal invariance broken by a mass term, the sound speed and bulk viscosity remain close to their 
conformal-limit values for a wide range of energies~\cite{Buchel10}.  

We emphasize that this question is distinct from that of small deviations from scale
invariance in weakly interacting $\2d$ Bose gases. Quantum gases with an unregularized delta-function interaction
have an SO(2,1) symmetry~\cite{Pitaevskii97} and exhibit scale invariance. 
However, the cutoff essential to describe an actual short-range interaction 
leads to a violation of scale invariance (analogous to an anomaly in quantum field theory) 
and an interaction-dependent shift in the breathing mode frequency from $2\omega_0$~\cite{Olshanii10}
in a $\2d$ Bose gas. The $\2d$ Bose gas experiments that see nearly scale-invariant behavior 
are in the weakly-interacting regime~\cite{Rath10, Hung11,cylinder}, where deviations are expected to be small.  
In contrast, the $\2d$ Fermi case that we focus on is not weakly interacting and we must 
take into account strong interactions.

{\textit{Results---}} We begin by summarizing our approach and main results. 
We consider a dilute Fermi gas in $d=2, 3$ dimensions with a short-range $s$-wave interaction,
arising from a broad Feshbach resonance between two spin species, each with density $n/2$. 
The dimensionless interaction $g_d$ is expressed as
$g_3 = -1/k_F a_3$ in $\3d$ and $g_2 = \log(k_F a_2)$ in $\2d$; $a_d$ is the $s$-wave scattering length that sets the dimer binding energy
$\varepsilon_b = -1/ma_d^2$. The fermions have mass $m$, density $n \sim k_F^d$, and we set $\hbar = 1$.

The quantity of central interest in our analysis is
\beq 
\gamma_d \equiv (1+2/d)P-\rho(\partial P/\partial\rho)_s, 
\label{gamma1}
\eeq
which is the deviation of the adiabatic compressibility $\rho(\partial P/\partial\rho)_s$ 
from its value $(1+2/d)P$ in a scale-invariant system, where the pressure $P\propto \rho^{(1+2/d)}$.
Here $s = S/N$ is the entropy per particle and $\rho = mn$ the mass density. 

\begin{center}
\begin{figure}
\includegraphics[width=0.33\textwidth]{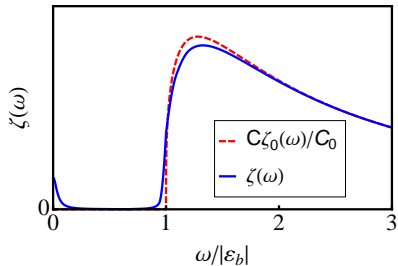}
\caption{
Schematic plot of the bulk viscosity spectral function $\zeta(\omega)$ (solid line) and the scaled ``vacuum" contribution $C\zeta_0(\omega)/C_0$ (dashed line) given by (\ref{zeta0}). The smallness of the subtracted sum rule (\ref{subtracted-sum-rule}) for $\left[ \zeta(\omega) - {C}\zeta_0(\omega)/ C_0 \right]$ 
implies very little spectral weight below the dimer binding energy $|\varepsilon_b|$ for excitations that break scale invariance.  
}
\label{zetafig}
\end{figure}
\end{center}

First, we show that $\gamma_d$ governs the difference between the frequency $\omega_m$ of the 
hydrodynamic monopole breathing mode and $2\omega_0$ for a Fermi gas 
in an isotropic harmonic trap $V_{\rm ext}(r) = m\omega_0^2r^2/2$. We find
\beq \
\omega_m^2/4\omega^2_0 = 1 -\left. \frac{d^2}{8}\int d^d\br\ \gamma_d(r) \right/ \int d^d\br\ n(r)V_{\mathrm{ext}}(r).
\label{monopole1}
\eeq
Second, we show that $\gamma_d$ is related to the exact sum rule
for the bulk viscosity spectral function $\zeta(\omega)$:
\beq
(2 / \pi)\int_0^\infty d\omega\left[ \zeta(\omega) - {C}\zeta_0(\omega)/ C_0 \right] = \gamma_d.
\label{subtracted-sum-rule}
\eeq
Here, $C$ is the ``contact''~\cite{Tan08a,Braaten08} with $\zeta_0(\omega)$ and $C_0$ the bulk viscosity
and contact in the zero-density $n \rightarrow 0$ limit at $T\!=\!0$.  The subtraction on the left-hand-side removes the large-$\omega$ tail of
$\zeta(\omega)$ (see Fig.~\ref{zetafig}) and the sum rule thus measures the availability of \emph{low-energy} ($\lesssim |\varepsilon_b|$) spectral weight for excitations that break scale invariance. 

The experimental observations of Ref.~\cite{Vogt12} in $\2d$ Fermi gases imply that $\gamma_2 \ll n\epsilon_F$.
Our goal is to understand \emph{why} $\gamma_2$ is so small for a wide range of interaction strengths, even though other measures of the departure from 
scale invariance, such as $P-2\varepsilon/d$ (where $\varepsilon$ is the energy density), are {\it not} small. 
$\gamma_d$ strictly vanishes only at the unitary point $g_3 = 0$ in $\3d$, and 
in the weak-coupling BCS limit  $g_2 \rightarrow \infty$ in $\2d$. However, we will argue that there is
considerable evidence for an anomalously small $\gamma_2$ across the entire BCS-BEC crossover.
Remarkably, within mean field theory~\cite{Randeria89} $\gamma_2^{\rm MF} = 0$ for all values of $g_2$ (and only in $d=2$).
In addition, the available $T\!=\!0$ quantum Monte Carlo (QMC) data in $\2d$~\cite{Bertaina11} leads to an estimate
of $\gamma_2$ that is consistent with zero over the entire crossover, except possibly
near $g_2 = 0$. We reach the same conclusion at finite $T$ using a scaling argument, and
 argue that this is due to the logarithmic dependence of $g_2$ on density.
Using the sum rule  (\ref{subtracted-sum-rule}), we will argue that a small $\gamma_2$ 
also gives insight into the negligible viscous damping of the monopole mode.

{\textit{Monopole breathing mode}---}
The  normal mode solutions of the hydrodynamic equations with frequency $\omega$ are
obtained from the Lagrangian~\cite{Taylor09}
\begin{align} {\cal{L}}[\bu] & =  \omega^2 \int d\br\;\rho_{0}\bu^2(\br)-\int \!\!d\br  \Big[\rho^{-1}_0\!\left(\partial
P/\partial\rho\right)_{\!s}(\delta\rho)^2 \nonumber\\& +  2\rho_0\!\left(\partial
T/\partial\rho\right)_{\!s}\delta\rho\delta s + \rho_0\left(\partial
T/\partial s \right)_{\!\rho}(\delta s)^2\Big], \label{TFVar}\end{align} 
describing quadratic fluctuations in entropy 
$\delta s$ and density $\delta\rho$ about their equilibrium values, $s_0$ and $\rho_0$.  The displacement field $\bu(\br,t)$ is related to the velocity $\bv$ by $\partial \bu/\partial t = \bv$.  Conservation of density and entropy gives $\delta\rho = -\bnab\cdot(\rho_{0}\bu)$ and $\delta s = -\bu \cdot\bnab s_0$.  
Eq.~(\ref{TFVar}) is valid in both the normal as well as the superfluid phase, where it describes first sound 
with $\bv_n =\bv_s$~\cite{Taylor09}.   

We obtain the result (\ref{monopole1}) for the breathing mode frequency using the scaling ansatz
$\bu(\br,t) = u\, \br\cos(\omega t)$ in (\ref{TFVar}), together with the Maxwell relation 
$(\partial P/\partial s)_{\rho} = \rho^2_0(\partial T/\partial \rho)_s$ and the equilibrium identities 
$\nabla P_0 = (\partial P/\partial\rho)_{ s}\nabla\rho_0 +(\partial P/\partial s)_{\rho}\nabla s_0=-n_0\nabla V_{\mathrm{ext}}$ 
for $V_{\mathrm{ext}} = m\omega^2_0r^2/2$, and 
$\nabla T_0 = (\partial T/\partial\rho)_{ s}\nabla\rho_0 +(\partial T/\partial s)_{\rho}\nabla s_0   = 0$. The above scaling ansatz provides a rigorous upper bound on the mode frequency~\cite{Taylor09}.  Generalizing the variational ansatz to
$\bu= \br\sum_{n=0}u_n r^{2n}\cos(\omega t)$, it is easy to show that the corrections to (\ref{monopole1}) are governed by 
higher powers of $\gamma_d$.   Thus,  $\gamma_d$ rigorously determines the 
deviation of the monopole frequency $\omega_m$ from $2\omega_0$. 

We next relate $\gamma_d$ to the contact $C$, 
given by $C = 2\pi m\atw (\partial \varepsilon/\atw)_s$ in $\2d$~\cite{Werner10,Valiente11} and 
$C = 4\pi m \ath^2  (\partial \varepsilon/\partial \ath)_s$ in $\3d$~\cite{Braaten08}. 
We find $\gamma_2 \!=\! -[C\! +\! \tfrac{\atw}{2}(\partial C/\partial \atw)_{\!s}]/4\pi m$ 
and $\gamma_3 = -\left[C + \ath \left(\partial C/\partial\ath\right)_s\right]/36\pi m \ath$.
This makes it clear that $\omega_m = 2\omega_0$ is strictly valid only for $a_2 \rightarrow \infty$, 
the BCS limit in $\2d$, where $C \rightarrow 0$, and at unitarity in $\3d$, where $|a_3| \rightarrow \infty$.
On the other hand, the breathing mode frequency (\ref{monopole1}) is very sensitive to $\gamma_d \neq 0$ in both $\2d$ and $\3d$.  
Using $\int d^d\br n V_{\mathrm{ext}}\sim{\cal{O}}(N\epsilon_F)$, we estimate that a value of $\gamma_d$ as small as $0.1n\epsilon_F$ would give rise to a 
$5\%$ shift in $\omega_m$.  The fact that no such shift is observed~\cite{Vogt12} in $\2d$ 
indicates that we must understand why $\gamma_2\ll n\epsilon_F$ for a wide range of $g_2$ and $T$.

{\textit{Viscosity sum rules---}}
The bulk viscosity $\zeta$ is the only transport coefficient that damps the scaling flow $\bu \propto \br$~\cite{LLFM}.  
To gain insight into why it is small in $\2d$, we derive a new bulk viscosity sum rule that relates 
$\gamma_2$ to the low-energy spectral weight for excitations that break scale-invariance symmetry.

The bulk viscosity spectral function $\zeta(\omega)$ is related by a Kubo formula 
to the transverse $\chi_T(\bq,\omega)$ and longitudinal $\chi_L(\bq,\omega)$ 
current correlators:
$\zeta(\omega) = \lim_{q\to 0}{m^2 \omega}\left[\mathrm{Im}\chi_L-(2-2/d)\mathrm{Im}\chi_T\right]/{q^2}$.
Generalizing Ref.~\cite{Taylor10} to arbitrary $d$, we obtain 
the exact sum rule 
\beq
\frac{2}{\pi}\!\int^{\infty}_0\! d\omega\zeta(\omega) = -\left(2-2/d\right)X_T + X_L - \rho c_s^2. \label{zetasumrule}
\eeq
Here, $X_{T(L)} = \lim_{q\to 0}{\langle[\hat{j}^{x}_{-\bq},[\hat{H},\hat{j}^{x}_{\bq}]]\rangle_{T(L)}}/{q^2}$ with
the current
$\hat{j}^x_{\bq} = \sum_{\bk\sigma}[(2\bk+\bq)_x/2m]\hat{c}^{\dagger}_{\bk\sigma}\hat{c}_{\bk+\bq\sigma}$.
The subscript ${T(L)}$ denotes the transverse (longitudinal) $q \to 0$ limit~\cite{limits}, and 
$c_s \equiv (\partial P/\partial\rho)^{1/2}_s$ is the adiabatic sound speed.   
Evaluating the commutators in (\ref{zetasumrule}) for an isotropic pair potential with range $r_0$, we find 
the $\2d$ result
$({2}/{\pi})\int d\omega \zeta(\omega)= 2\varepsilon - \rho c^2_s + \alpha C/m + \beta C\ln\Lambda/m$.
Here, $\Lambda=1/ r_0$ is an ultraviolet (UV) cutoff, 
and $\alpha,\beta$ are constants.
In $\3d$~\cite{Taylor10}, the  terms proportional to $C$ are of the form $\alpha C/ma_3 + \beta C\Lambda/m$.

The key insight that allows us to obtain physical results independent of $\Lambda$ is that an UV divergence
of precisely the same form must arise in the two-body problem.  
The sum rule for $\zeta_0$  has the same form as above, but with energy density and contact replaced by their zero-density, 
$T\!=\!0$ values, $\varepsilon_0$ and $C_0$, while $c_s=0$ for $n \to 0$. The exact solution $\zeta_0$ of the two-body problem
can then be used to regularize the divergence in the many-body problem. 
The same idea underlies the standard replacement of the bare interaction with the two-particle $s$-wave scattering length 
in the study of dilute gases~\cite{BCSBEC}. 

The $T=0$, zero-density limit $\zeta_0(\omega)$ of the viscosity spectral function has an {\it exact} representation in terms of the sum of all
particle-particle ladder diagrams with two external current vertices. These are the well-known~\cite{LarkinVarlamaov}
 Aslamazov--Larkin, Maki--Thompson, and self-energy diagrams.  We thus obtain the $\2d$ result~\cite{Supplemental}
\beq 
\zeta_0(\omega) = \frac{C_0}{4m\omega}\frac{\Theta(\omega-|\varepsilon_b|)}{\ln^2(\omega/|\varepsilon_b|-1)+\pi^2}\label{zeta0}
\eeq
for $\omega>0$, and $\zeta_0(-\omega)=\zeta_0(\omega)$.  
In $\2d$, there is a bound state for all values of the scattering length~\cite{Randeria89}. Thus, in the zero-density (single dimer) and temperature limit, 
$\varepsilon_0 = \varepsilon_b= - 1/ma_2^2$ and $C_0= 4\pi/\atw^2$.  The absence of spectral weight in  $\zeta_0$  below $|\varepsilon_b|$  
is due to the fact that the only excitations in this limit involve pair disassociation with a 
gap $|\varepsilon_b|$ at $T=0$.

The UV divergences can now be removed by looking at the difference between the sum rule for the interacting many-body system
and that for the  $T=0$, $n \rightarrow 0$ limit, scaled by $(C/C_0)$.
In $\2d$, we find 
$({2}/{\pi})\int d\omega \left[\zeta(\omega)\!-\! C\zeta_0(\omega)/C_0\right] \!=\! 2\varepsilon - \rho c^2_s - 2{C}\varepsilon_0/C_0$.
Using $C = 4\pi m(P-\varepsilon)$ and $\varepsilon_0/C_0 = -1/(4\pi m)$,
we obtain (\ref{subtracted-sum-rule}) in $\2d$.
The same methodology can be also be used to obtain corresponding results for the bulk viscosity in $\3d$ as well
as the shear viscosity in any $d$~\cite{Supplemental}.

Our main focus will be on (\ref{subtracted-sum-rule}), which quantifies the
{\it low-energy} spectral weight in $\zeta(\omega)$ with the high-energy tail $C\zeta_0(\omega)/C_0$ subtracted out
; see Fig.~\ref{zetafig}.
However, we can also obtain the total spectral weight in $\zeta(\omega)$: 
\beq
S_{\2d} \!\equiv\! \frac{2}{\pi}\!\int_0^\infty\!\!d\omega \zeta(\omega) \!=\! 3P -\varepsilon- \rho c_s^2 = - \frac{1}{8\pi m}\left(\frac{\partial C}{ \partial g_2} \right)_{\!s}.
\label{zetasumrule2}
\eeq
$S_{\2d} \geq 0$
for all $g_2$~\cite{Werner10}, as required by $\zeta(\omega)\geq 0\;\forall \omega$~\cite{Taylor10}.  
\begin{figure}
\includegraphics[width=0.32 \textwidth]{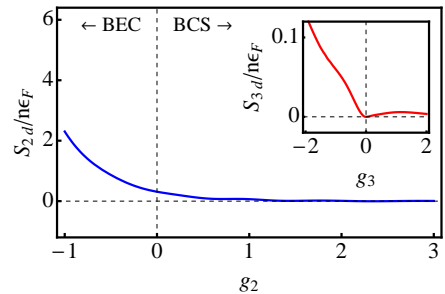}
\caption{\textit{Bulk viscosity sum rule:}   The $\2d$ sum rule $S_{\2d}$ at $T = 0$ in units of $n\epsilon_F$ plotted as a function 
of the coupling $g_2 = \gtw$.
Inset: The corresponding $\3d$ result~\cite{Taylor10} as a function of $g_3 = \gth$. In both $\2d$ and $\3d$, $g_d \to +\infty$ 
the BCS limit while $g_d \to -\infty$ is the BEC limit. 
}
\label{sumrulefig}
\end{figure}
In Fig.~\ref{sumrulefig}, we plot  $S_{\2d}$ as a function of $g_2 = \log(k_F a_2)$
using $T=0$ QMC data~\cite{Bertaina11} to evaluate the right-hand-side of (\ref{zetasumrule2}).
Both $S_{\2d}$ and its $\3d$ counterpart $S_{\3d}$~\cite{Taylor10}  (inset of Fig.~\ref{sumrulefig} using the QMC data of Ref.~\cite{Astrakharchik04}) 
are $\ll n\epsilon_F$ in the BCS region $g_d\gtrsim 1$, but become significantly larger on the BEC side.

{\textit{Apparent scale invariance}---}We now have all the results in hand to discuss deviations from
scale invariance. So far, we have shown that $\gamma_d$ controls the deviation of the monopole $\omega_m$ from $2\omega_0$ and also governs the availability of low-energy bulk viscosity spectral weight $\zeta(\omega)$.  
We can intuitively understand the exact relation (\ref{subtracted-sum-rule}) between a $\zeta$-sum rule and the shift in the mode
frequency as a Kramers--Kronig transform of the d.c.~bulk viscosity $\zeta(0)$ that damps 
the monopole mode.

What, if anything, is special about $\2d$ that leads to the strong experimental signatures~\cite{Vogt12} of scale invariance?  
We begin by addressing this question at $T=0$ and then generalize to finite temperatures.
The first clue comes from mean field theory (MFT),  which in $\2d$ has a transparent solution~\cite{Randeria89} 
across the entire $T=0$ BCS-BEC crossover:  $\varepsilon = n\epsilon_F/2 - n|\varepsilon_b|/2$.  This leads to $P = n\epsilon_F/2$ and thus
$\gamma_2^{\rm MFT} \equiv 0$ for {\it all} couplings $g_2$. Contrast this with the $\2d$ MFT result
$P-\varepsilon = n|\varepsilon_b|/2$, which is very small in the BCS regime
but very large on the BEC side. This is our first hint of something we will see again: $\gamma_d$ is small in part
because it does not involve physics on the scale of the dimer binding energy, whereas $P-\varepsilon$ does.

\begin{figure}
\includegraphics[width=0.34 \textwidth]{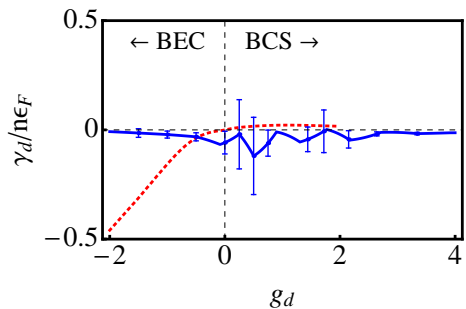}
\caption{$\gamma_{2}$ (solid blue line) and $\gamma_{3}$ (dashed red line) shown in units of $n\epsilon_F$, where $n$ and $\epsilon_F$ are the two and three dimensional density and Fermi energy, respectively.  The error bars on $\gamma_2$ are associated with numerical derivatives of QMC data 
(with errors)~\cite{Bertaina11}.  The coupling $g_d$ is $\gtw$ for $d= 2$ and $\gth$ for $d=3$.}
\label{Subtractedfig}
\end{figure}

To understand how quantum fluctuations beyond MFT affect the result for $\gamma_2$, we use
$T\!=\!0$ QMC~\cite{Bertaina11}. 
We find that the QMC-derived $\gamma_{2}$
is vanishingly small in both BCS and BEC regimes, and even for $g_2\sim 0$, $\gamma_{2}\sim 0$
(within large error bars) as shown in Fig.~\ref{Subtractedfig}.  We also see from
this figure that the $\2d$ result is quite different from the $\3d$ case. The QMC estimate for $\gamma_3$ (using data from Ref.~\cite{Astrakharchik04}), though quite small on the BCS side of the crossover, is large in the BEC region in $\3d$.

We now show that the difference between $\2d$ and $\3d$ is tied to the
form of the dimensionless couplings $g_2 = \log(k_F a_2)$ and $g_3 = -1/k_F a_3$.
In the BCS limit ($g_d \gg 1$), the equation of state has the form
$\varepsilon = (n\epsilon_F/2)[1 + A/g_d + B/g_d^2 + \ldots]$ in both $\2d$ and $\3d$.
The perturbative Hartree plus ``Fermi liquid'' corrections are larger than the 
pairing contribution not shown. (The only qualitative difference is that A is negative in $\3d$
but positive in $\2d$~\cite{Engelbrecht92}.)
Both $\gamma_d$, calculated using (\ref{gamma1}), and $P-\varepsilon = (a_d/d)\left(\partial\varepsilon/\partial a_d\right)$,
are small in the BCS limit. In the BEC limit ($g_d < 0$ and $|g_d| \gg 1$), we get $\varepsilon =  - n|\varepsilon_b|/2 + \ldots$,
which is the energy density of $n/2$ dimers with perturbative corrections in powers of $1/|g_d|$. The key difference between
$\2d$ and $\3d$ is in the $g_d$-dependence of the binding energy $|\varepsilon_b|$, 
which $\sim \exp(|g_2|)$ in $\2d$ and $\sim 1/|g_3|^2$ in $\3d$.

To understand the effects of finite temperature, we write
the pressure and energy density, related by $P = n(\partial \varepsilon/\partial n)_s - \varepsilon$,
in the scaling forms $P = n\epsilon_F{\cal F}(g_d,s)$ and $\varepsilon = n\epsilon_F{\cal E}(g_d,s)$. 
There is a qualitative difference between the $g_2$-dependence
of the scaling functions ${\cal F}$ and ${\cal E}$ in $\2d$. The pressure does not have a
contribution on the scale of the dimer binding energy $|\varepsilon_b| = 1/ma_2^2$; i.e.,
it does not have a potentially exponentially large contribution in $g_2 = \log(k_F a_2)$,
while the energy density does. We have already seen this in the $T\!=\!0$ MFT results,
and the same is also observed in the $\2d$ virial expansion~\cite{Taylor13}. 
We conjecture that the scaling function ${\cal F}$ is a slowly varying function of $g_2$ at all temperatures in $\2d$ 
(except in the immediate vicinity of a weak singularity at $T_c$).
The equation of state is then $P \sim n^2$ up to logarithmic corrections, leading to a small
$\gamma_2$.  

The absence of high energy contributions on the scale of the dimer binding energy
to $P$ and the compressibility is also consistent with $\gamma_2$ being related
to the low energy spectral weight as shown by our sum rule.
Once high energy excitations on the scale of $|\varepsilon_b|$ are excluded,
low energy phonons (with a near scale-invariant dispersion $\omega_{\bq}(n) \sim \sqrt{n}q$), for instance,  dominate the equation of state leading to $P \sim n^2$
and a small $\gamma_2$. Unlike $\gamma_2$, however, $P-\varepsilon$ is not small, 
as it involves high energy contributions on the scale of $|\varepsilon_b|$ in the BEC regime.

Another way to characterize the deviation from scale invariance, analogous to the
``trace anomaly'' in quantum field theory, is to rewrite (\ref{gamma1}) as
as $\gamma_d = - (\partial P / \partial{g_d}) \beta(g_d)/d$, where $\beta(g_d) \equiv k_F(\partial g_d/\partial {k_F})$  describes the scaling of 
the coupling $g_d$ with respect to the momentum scale $k_F$.  
We see that $\beta(g_2)=1$ while $\beta(g_3)=g_3$, reflecting the difference between the
logarithmic and power-law dependence on the density in $\2d$ and $\3d$ respectively.  
In both the BCS ($g_d\gg 1$) and BEC ($g_d\ll -1$) regions, the $\2d$ beta function is much smaller than its $\3d$ counterpart.  

Finally, using the sum rules  (\ref{zetasumrule2}) and (\ref{subtracted-sum-rule}), 
we discuss the damping of the monopole mode, controlled by $\zeta(0)$.  
Although a small value for the sum rule by itself does not rigorously
upper-bound $\zeta$, any physically reasonable functional form for the spectral function 
(e.g., a Drude form for $\omega\lesssim |\varepsilon_b|$; see Fig.~\ref{zetafig}) 
would lead to a very small value for $\zeta(0)$.
We see from Fig.~\ref{sumrulefig} 
that in the BCS regime $g_d \gtrsim 1$, the sum rule $\int d\omega \zeta(\omega) \ll n\epsilon_F$
in both $\2d$ and in $\3d$. We would thus expect a very small $\zeta \ll n$ here in both
$\2d$ and $\3d$. This sum rule is quite large on the BEC side and does not lead to any
restriction on $\zeta$. From the low-energy sum rule (\ref{subtracted-sum-rule}), however, we see 
that the large value of $S_{2d}$ in the BEC limit is entirely dominated by the high-energy
tail on scales larger than the dimer binding energy. Once this is subtracted out, the low-energy
integrated spectral weight, equal to $\gamma_2$, is very small even in the
BEC regime (see Fig.~\ref{zetafig}). Thus in $\2d$, we expect the bulk viscosity $\zeta$ to be very small both in the 
BCS and in the BEC regimes.

{\textit{Conclusions}}---We have shown that the parameter $\gamma_d$ controls
the deviation of the breathing mode frequency $\omega_m$ from its scale-invariant value $2\omega_0$ and 
also quantifies the low-energy spectral weight for excitations that break scale invariance,
using an exact sum rule. We argue that $\2d$ is special,
with a coupling that depends logarithmically on density, leading to a very small $\gamma_2$,
 even in the BEC regime where scale invariance is strongly broken 
by the large dimer binding energy (and hence $P\!-\!\varepsilon$ is large). 
The small $\gamma_2$ also implies, via the $\2d$ sum rule, weak damping of the monopole mode in
both the BCS and BEC regimes.  The regime very near $g_2=0$ deserves further theoretical and experimental
investigation, but the available evidence suggests that $\gamma_2$
might be small there as well.
     
{\textit{Acknowledgments}}---MR acknowledges support from the NSF grant DMR-1006532, and ET from
NSERC and the Canadian Institute for Advanced Research (CIFAR).  

{\textit{Note added in proof}}---While completing this manuscript we became aware of Ref.~\cite{Hofmann12}, which analyzes the experiment
of Ref.~\cite{Vogt12} from the different perspective of quantum anomalies at $T\!=\!0$.  In the only area of substantial overlap,
our general result (\ref{monopole1}) for the breathing mode reduces to that of Ref.~\cite{Hofmann12} if we assume a polytropic equation of state.


\clearpage

\section{Supplemental Material}

\begin{figure}
\includegraphics[width=0.4 \textwidth]{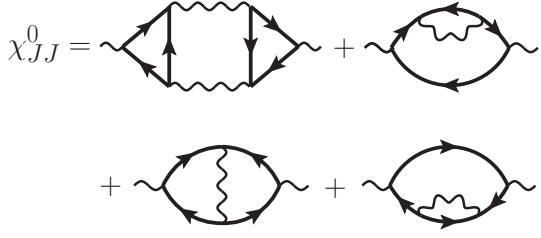}
\caption{Exact zero-density limit of the current correlation function in $\2d$.  Clockwise from top-left: Aslamazov--Larkin, the two self-energy, and Maki--Thompson diagrams. 
Lines with arrows denote the vacuum single-particle Green's function $G$; wavy lines, the pair propagator $\Gamma$. The external vertices are the bare current vertices $\gamma_x$.}
\label{diagrams1}
\end{figure}

{\textit{Viscosity in the zero-density limit---}}In the zero-density limit, the current correlation function can be calculated without approximation by summing up all ladder diagrams.  (A related approximate calculation for the finite density viscosity spectral functions was carried out in Ref.~\cite{Enss11s}.) It receives contributions from  Aslamazov-Larkin (AL), Maki-Thompson (MT), and self-energy (SE) diagrams, as shown in Fig.~\ref{diagrams1} (as long as the energy of two particles in vacuum is nonzero and negative, as expected in two dimensions~\cite{Randeria89s}, the bare loop contribution vanishes). Here, the bold lines are zero-density Green's functions, $G(k) = (ik_n-\xi_{\bk})^{-1}$, with $\xi_{\bk} = \epsilon_{\bk}-\mu$, $\epsilon_{\bk}=\bk^2/2m$, and $\mu = -|\varepsilon_b|/2+ 0^+$. $\hat{j}^x_{\bq} = \sum_{\bk\sigma}\gamma_x(\bk,\bk+\bq)\hat{c}^{\dagger}_{\bk\sigma}\hat{c}_{\bk+\bq\sigma}$ 
is the $x$-component of the current operator, with $\gamma_x(\bk,\bk^{\prime})\equiv (\bk+\bk^{\prime})_x/2m$ the bare current vertex.  
The internal wavy lines are the vacuum pair propagator $\Gamma$, the inverse of which is given by  
\beq
\Gamma^{-1}(q)= -\frac{1}{g} + \frac{1}{\beta}\sum_kG(-k)G(k+q)\label{Gamma0}\eeq
Using the renormalization scheme~\cite{Randeria89s}
\beq g^{-1} = T^{-1}(E) - \sum_{\bk}(2\epsilon_{\bk}-2E)^{-1},\eeq
where $E$ is some arbitrary (nonzero) energy and
\beq T^{-1}(\omega) =  \frac{m}{4\pi}\left[\ln(|\varepsilon_b|/\omega)+i\pi\right],\label{T2D}\eeq  
in (\ref{Gamma0}) gives the convergent result (after analytic continuation of the Bose frequency $iq_m\to \omega+i0^+$)
\beq \Gamma^{-1}(\bq,\omega) = \frac{m}{4\pi}\ln\left[1-(\omega+i0^+-\bq^2/4m + \delta\mu)/|\varepsilon_b|\right].\label{Gamma1}\eeq
Here, we have defined $\mu = -|\varepsilon_b|/2 + \delta\mu$.  $\delta\mu$ approaches zero from above in the zero-density limit.  

Considering the correlation function for the $x$-component $j^x$ of the current, the contributions shown in Fig.~\ref{diagrams1} are given explicitly below.  The Aslamazov-Larkin contribution is 
\beq \chi_{\mathrm{AL}}(q) = -\frac{1}{\beta}\sum_{p}S^2_x(p,q)\Gamma(p)\Gamma(p+q), \label{AL}\eeq
with the block $S_x(p,q)$ of three Green's functions given by
\beq S_x(p,q) = \frac{2}{\beta}\sum_{k} \gamma_x(\bk,\bk+\bq)G(k+q)G(k)G(p-k).\label{Sblock}\eeq
The two self-energy contributions are
\begin{align}\chi_{\mathrm{S}}(q) &= -\frac{2}{\beta^2}\sum_{p,k}\gamma^2_x(\bk,\bk+\bq)\Gamma(p)G^2(k)G(p-k)\times\nonumber\\& G(k+q) + (q\to -q)\label{S}\end{align} 
and the Maki-Thompson contribution is
\begin{align} \chi_{\mathrm{MT}}(q) &= -\frac{2}{\beta^2}\sum_{p,k}\gamma_x(\bk,\bk+\bq)\gamma_x(\bp-\bk,\bp-\bk-\bq)\times\nonumber\\&\Gamma(p)G(k)G(p-k)G(k+q)G(p-k-q).\label{MT}\end{align}
$q\equiv (\bq,iq_m)$, $p\equiv (\bp,ip_m)$, and $k \equiv (\bk,ik_n)$ are 4-vectors, where $q_m,p_m$ and $k_n$ are Bose and Fermi Matsubara frequencies, respectively. 
The pre-factors of $2$ appearing in (\ref{Sblock}), (\ref{S}), and (\ref{MT}) arise since there are two spin degrees of freedom. 

Using (\ref{Gamma1}), the contributions to the density correlation are found after some straightforward, if modestly laborious, algebra.  To order $\bq^2$, the transverse and longitudinal components of the MT + SE contributions are
\beq \left.\chi^{\prime\prime}_T(\bq,\omega)\right|_{\mathrm{MT}+\mathrm{SE}} = \frac{C_0 q^2f(\omega)}{8m(m\omega)^2}\Theta(\omega-|\varepsilon_b|)-(\omega\to -\omega),\eeq
\beq \left.\chi^{\prime\prime}_L(\bq,\omega)\right|_{\mathrm{MT}+\mathrm{SE}} = \frac{3C_0 q^2f(\omega)}{8m(m\omega)^2}\Theta(\omega-|\varepsilon_b|)-(\omega\to -\omega).\eeq
As in $\3d$, the AL contribution is purely longitudinal~\cite{Enss11s}:
\begin{align}& \left.\chi^{\prime\prime}_L(\bq,\omega)\right|_{\mathrm{AL}} = \frac{C_0 q^2}{4m(m\omega)^2}\frac{\Theta(\omega-|\varepsilon_b|)}{g(\omega)+\pi^2}\times \nonumber \\ &\;\;\;\left[1-(g(\omega)+\pi^2)f(\omega)\right]-(\omega\to -\omega).\end{align}
In the above, we have defined $f(\omega)\equiv (1-|\varepsilon_b|/\omega)^2$ and $g(\omega)\equiv \ln^2 (\omega/|\varepsilon_b|-1)$.  $C_0\equiv 4\pi m|\varepsilon_b|$ is the zero-density limit of the contact.  
Using these in the appropriate Kubo formulae (see main text, above  (\ref{zetasumrule})), one finds (\ref{zeta0}).
Applying this procedure to the $\3d$ case, one reproduces the $\zeta(\omega)$ sum rule reported in 
Ref.~\cite{Taylor10s}.

{\textit{Shear Viscosity ---}} In the paper we focus on the bulk viscosity sum rule, since this $\zeta$ is relevant for the 
damping of the monopole breathing mode and has direct bearing on the question of scale invariance.
For completeness, we briefly discuss here the shear viscosity sum rule.

We first use the above results to evaluate the zero-density, $T=0$ limit $\eta_0(\omega)$ of the shear viscosity
$\eta(\omega) = \lim_{q\to 0}{m^2 \omega}\mathrm{Im}\chi_T/{q^2}$, with
$(2/\pi)\!\!\int\! d\omega \eta(\omega)\!=\!X_T$.  For the even spectral function [$\eta_0(-\omega)=\eta_0(\omega)$], we find
\beq \eta_0(\omega>0) = ({C_0}/{8m\omega}) (1 - |\varepsilon_b|/\omega)^2 \Theta(\omega-|\varepsilon_b|),\eeq consistent with the high-frequency spectral tail calculated in Ref.~\cite{Hofmann11s}.

We next write down the convergent subtracted sum rule for the shear viscosity.  Evaluating the commutators in (5) in the main text, one finds $({2}/{\pi})\int d\omega \eta(\omega)=\varepsilon + \alpha' C/m + \beta' C\ln\Lambda/m$ for the shear viscosity sum rule.  As with the bulk viscosity sum rule, $\Lambda=1/ r_0$ is an ultraviolet (UV) cutoff, and $\alpha',\beta'$ are constants.  Subtracting the sum rule for $C\eta_0(\omega)/C_0$ from the sum rule for the full many-body viscosity, one finds
\begin{align}(2\pi)\int^{\infty}_0 d\omega \left[\eta(\omega)\!-\! C\eta_0(\omega)/C_0\right] =& \varepsilon - {C}\varepsilon_0/C_0\nonumber\\ =&\varepsilon + C/4\pi m.\end{align}
Following the same procedure to obtain a subtracted shear viscosity sum rule in $3d$ gives the sum rule obtained in Ref.~\cite{Enss11s}.

\end{document}